**Four Wave Mixing Spectroscopy at the Interface Between the Time and Frequency Domains**

Yuri Paskover[*], Andrey Shalit[*] and Yehiam Prior
Department of Chemical Physics
Weizmann Institute of Science, Rehovot, Israel 76100

Combined spectrally and time resolved measurements provide information not otherwise available if performed in either domain alone. We demonstrate a new approach to Four Wave Mixing spectroscopy, where spectral selectivity is achieved by phase matching filtering without a spectrometer, and the time resolved signal is obtained within a single pulse and without mechanically scanning any delays. We analyze the Degenerate Four Wave Mixing signal, and show that a counter-rotating Feynman diagram not previously considered is necessary in order to understand the measured frequency and time resolved spectrograms.

Spectrally and temporally resolved measurements are considered to be complementary, and the choice of one or the other is usually based on practical considerations such as signal to noise ratios of specific experiments. Coherent spectroscopy with ultrashort pulses enables direct observation of specific transitions, as well as real time tracking of intramolecular pathways of coherence or energy redistribution. The inherently broad bandwidth of the ultrashort pulses provides an opportunity for simultaneous excitation of multiple degrees of freedom and for measurement of their interactions[1], as in multidimensional NMR[2,3] .

Typically, time resolved probing of molecular dynamics involves a Four Wave Mixing (FWM) experiment where initially two photons are used to excite a vibrational/rotational wavepacket via a stimulated Raman type process, and a third, time delayed pulse, stimulates coherent emission of the FWM signal in the phase matching direction[4]. With the extremely broad bandwidth of ultrashort pulses, the traditional classification of such experiments to CARS (Coherent Anti-Stokes Raman Scattering), CSRS (Coherent Stokes Raman Scattering) or DFWM (Degenerate Four

Wave Mixing) might be misleading, and therefore we will use the generic, more accurate term Four Wave Mixing (FWM).

Combined time and frequency domain measurements of coherent molecular dynamics were demonstrated to provide information not otherwise available in either domain, simplifying the interpretation of the retrieved data. Heid et al. reported on spectrally dispersed femtosecond time-resolved CARS for the investigation of the electronic ground-state vibrational dynamics of complex molecules in solution[5]. They showed that the two-dimensional mapping allows for interpretation of the complex signal produced by molecules with a large number of molecular vibrations. Prince et al. suggested a variant of time resolved CARS that allows for multiplex detection of Raman active modes on the ground and excited electronic states [6]. More recently, Nath et al. [7,8] showed that for CARS, simultaneous detection in the time and frequency domains provides information even when the pulse durations and spectral widths do not allow either full temporal or full spectral resolution. Moreover, these authors have shown that from the combined time frequency domain (TFD) measurements one can derive the full (amplitude and phase) information on the third order susceptibility $\chi(3)$. Similar results were recently demonstrated by Konorov et al. [9,] and Xu et al. [10] who have shown that an arrangement similar to XFROG is useful for the extraction of full characterization of the third order susceptibility from TFD data.

In this paper we reiterate the previous results, we discuss another situation where the TFD is a very useful regime to work in as it provides additional information not available in either domain alone. We show that when low frequency modes are measured in a Degenerate Four Wave Mixing (DFWM) experiment where all frequencies are the same, an additional double sided Feynman diagram not previously considered provides enough information to fully identify and separate fundamental and beat modes observed in the spectrum. We go a step further and demonstrate how the full TFD information can be derived with only a small number of laser pulses. The structure of the paper is as follows: We first discuss TFD measurement of chloroform obtained in the standard manner, where the generated FWM signal is frequency dispersed by a monochromator and measured on a CCD. We provide a detailed analysis of the coherent pathways (Feynman diagrams) leading to the generation of this signal, and show how the TFD information is essential for the characterization of the observed data, and how without it, fundamental or beat modes cannot be identified. Next we introduce a new experimental approach to the observation of the TFD in a single shot. The new method is an expansion of our previously demonstrated single shot CARS measurement and a newly introduced Phase Matched Filtering, which provides for the recording of the full TFD data without scanning any delays and without a spectrometer. We conclude the paper with a discussion of the advantages and potential of the new approach as well as its possible limitations.

Consider a time resolved FWM experiment in neat liquid chloroform. The three input pulses are derived from the same regeneratively amplified laser (60fs, 800 nm central frequency, ~1 mJ per pulse, 1 KHz repetition rate). The experimental geometry is the now standard three dimensional folded Boxcars phase matching arrangement[11] which enables the spatial separation of the signal beam. The generated FWM signal is spectrally dispersed in a spectrometer before being detected on a CCD. The first two pulses create a ground state coherent superposition of rovibrational states, from which a time delayed probe pulse is being scattered to generate the FWM signal. Figure 1(a) depicts a spectrogram of such measurements: for each delay of the probe pulse (horizontal axis) the FWM signal is spectrally resolved on the spectrometer (vertical axis).  In this spectrogram, the vertical axis is calibrated (cm$^{-1}$) in terms of the frequency detuning of the detected signal from the laser center frequency (around 800 nm).  The horizontal axis is the probe delay mechanically scanned over the range of 500-2500 fs with a 20 fs step size.  The measurement was intentionally started at $\tau = 500\,fs$ to avoid the very strong coherence peak.

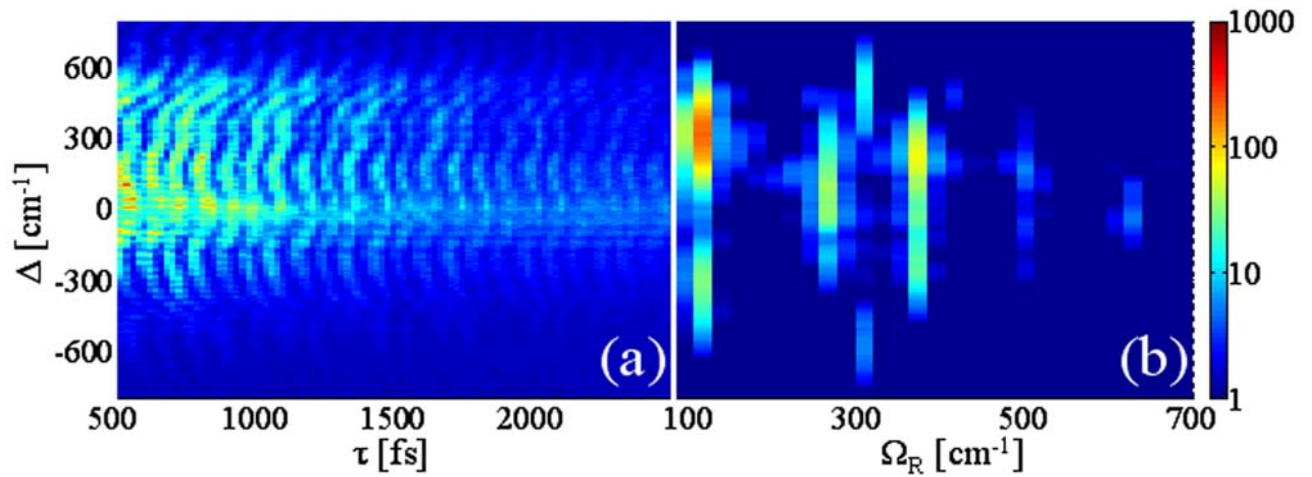

FIG. 1. (Color online) (a) Spectrally and temporally resolved FWM signal from neat chloroform (CHCl$_3$). (b) Horizontal line by line Fourier Transform of part (a) depicting the power spectrum at each Raman frequency (log scale).

Figure 1(b) is a two dimensional plot consisting of a composite horizontal line by line Fourier Transform of the data in Fig. 1(a). Thus, the horizontal axis is now converted to the frequency domain, showing the spectral contents (vertical axis) of the generated FWM signal at each frequency (horizontal axis). In Fig. 1(b), there are 5 main features observed around 104, 261, 300, 365 and 625 cm$^{-1}$. These are fundamental vibrational modes as well as beat frequencies between them.   As noted above, these measurements of low frequency Raman modes are similar and complementary to the measurements by Nath et al.[8] and Urbanek et al.[12]  who measured higher frequencies in non degenerate CARS experiments, and have shown that the full complex information on the susceptibility can be derived from the TFD data.

Even a simple, qualitative inspection of Fig. 1(b) reveals differences between the observed signals at different Raman frequencies. Thus, the signal at 260 cm$^{-1}$ is centered around zero detuning from the laser frequency, while the signal at 104 cm$^{-1}$ appears to be split, peaking around $\pm$ 310 cm$^{-1}$. This additional information, not available in either the time or frequency domains alone, is very useful for the characterization of the observed modes, as is discussed theoretically in the next section.

However, already at this point it should be noted that the acquisition of a signal of the type depicted in Fig. 1 takes a long time, as the delay needs to be scanned mechanically, and for each delay, the spectrum is acquired with proper signal averaging. Long measurement times may pose a problem for molecules which are not stable against photobleaching, and in particular in cases of resonant excitation. Thus, a significant advantage will be added when the full signal can be captured on a much faster time scale, as is discussed below.

In this section we consider Time Resolved FWM (TRFWM) in which all input pulses are derived form the same laser and have the same spectral properties. All Raman type, two photon transitions are driven form the spectral components included within the bandwidth of the pulses. The observed the signal is produced when the probe pulse, being scattered off by the polarization induced in the medium by the first two pulses, generates a third order polarization.

Perturbation theory provides a formal description for this nonlinear polarization:

$$P^{(3)}(t) = \left\langle \Psi^{(0)}(t) \middle| \hat{\mu} \middle| \Psi^{(3)}(t) \right\rangle + \left\langle \Psi^{(1)}(t) \middle| \hat{\mu} \middle| \Psi^{(2)}(t) \right\rangle + c.c. \qquad (1)$$

Here the $P^{(3)}(t)$ is the third order polarization, $\Psi^{(n)}(t)$ is wavefunction representing the state of the molecule after interaction with $n$ fields, and $\hat{\mu}$ is the transitional dipole operator (assumed, for simplicity, to be the same for all transitions).

As discussed many times in the past there are many coherent pathways leading to the third order susceptibility, involving the same input frequencies, but differing in their time ordering. These pathways are conveniently represented by double-sided Feynman diagrams[13]. For the configuration discussed here of a nonresonant TRFWM in a three dimensional boxcars configuration, there are two dominant diagrams, these are given in Table :

**Table I.** The two leading processes contributing to third order polarization in non-resonant time delayed FWM in 3-D folded Boxcars configuration. The left column depicts the double sided Feynman diagrams while right column sketches the corresponding energy level diagram. The solid and dashed lines symbolize the "ket" and the "bra" side transitions respectively.

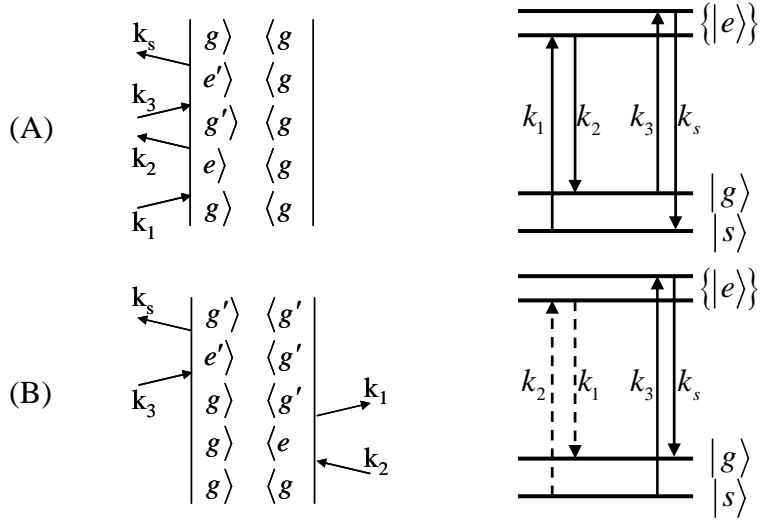

The input laser fields are assumed to be transform limited pulses, whose electric field is given by:

$$E_n(t) = \varepsilon_n(t)e^{-i\omega_n t} + c.c. = \frac{1}{\sqrt{2\pi}}\int\limits_{-\infty}^{\infty}d\omega S_n(\omega)e^{-i\omega t} + c.c.$$

Here $E_n(t)$ is the electric field of the pulse, $\varepsilon_n(t)$ is the pulse envelope, $\omega_n$ is its carrier frequency, $S_n(\omega)$ is the spectral amplitude of the $n$-th pulse (here the same for all pulses)

To lowest perturbation order the expressions for the third order polarization for each of the diagrams (A,B) is given by:

$$P_A^{(3)}(t,\tau) = i\sum_g \frac{\left|\langle s|\hat{\alpha}_g|g\rangle\right|^2}{2\pi\Delta^2\hbar^3}\varepsilon_3(t)e^{-i\left[\left(\omega_3 + \Omega_g\right)t + \Omega_g\tau\right]}R_{1,2}\left(\Omega_g\right) + c.c. \qquad (2)$$

$$P_B^{(3)}(t,\tau) = -i\sum_g \frac{\left|\left\langle s\left|\hat{\alpha}_g\right|g\right\rangle\right|^2}{2\pi\Delta^2\hbar^3}\varepsilon_3(t)e^{-i\left[(\omega_3-\Omega_g)t-\Omega_g\tau\right]}R_{2,1}^*\left(\Omega_g\right) + c.c. \qquad (3)$$

Here $\tau$ represents the probe delay, $\Delta$ is the nominal detuning of the laser frequency from the nearest electronic transition, the index $g$ accounts for all allowed vibrational transition frequency $\Omega_g$ and polarizability $\hat{\alpha}_g$. Note, that for a set of input pulses, the bandwidth available for excitation at of frequency $\Omega$ is given by a combination of frequency components derived from the input pulses:

$$R_{m,n}(\Omega) = \int_{-\infty}^{\infty}d\omega S_m(\omega)S_n^*(\omega-\Omega).$$

In a recent paper we have analyzed in detail the generation of a Time Resolved Four Wave Mixing (TRFWM) signal, and have shown that unless the FWM signal is linearized, the fundamental vibrational frequencies are not found in the signal and only combination frequencies are observed. We have further shown that the presence of slow rotational excitation in the system[14] gives rise to in-situ heterodyning detection, thus linearizing the signal. Taking an additional slow rotational contribution the total polarization of the medium can be written as

$$P_{Tot}^{(3)}(t,\tau) = P_{rot}^{(3)}(t) + \sum_g\left[P_A^{(3)}\left(t,\Omega_g,\tau\right) + P_B^{(3)}\left(t,\Omega_g,\tau\right)\right].$$

Because of its slow response, the rotational polarization is independent of the probe delay. The spectral properties of the field produced by each one of the components of the polarization may be obtained by its direct transform from time to frequency domain.

$$
\begin{aligned}
P_{rot}^{(3)}(\omega) &\propto S_3(\omega)\\
P_A^{(3)}(\omega,\Omega_g,\tau) &\propto S_3\left(\omega-\Omega_g\right)e^{-i\Omega_g\tau} + c.c.\\
P_B^{(3)}(\omega,\Omega_g,\tau) &\propto S_3\left(\omega+\Omega_g\right)e^{i\Omega_g\tau} + c.c.
\end{aligned}
\qquad (4)
$$

The detected quantity is the intensity of the emitted signal, and thus proportional to the modulus square of the total polarization.

$$
\begin{aligned}
I\left(\omega,\tau\right) &\propto \left|P_{Tot}^{(3)}\left(\omega,\tau\right)\right|^{2} \\
&= \left|P_{rot}^{(3)}\left(\omega,\tau\right)\right|^{2} \\
&+ \sum_{g,g'}\left\{P_{A}^{(3)}\left(\omega,\Omega_{g},\tau\right)\left[P_{A}^{(3)}\left(\omega,\Omega_{g'},\tau\right)\right]^{*} + P_{B}^{(3)}\left(\omega,\Omega_{g},\tau\right)\left[P_{B}^{(3)}\left(\omega,\Omega_{g'},\tau\right)\right]^{*}\right\} \\
&+ \sum_{g}\left\{P_{A}^{(3)}\left(\omega,\Omega_{g},\tau\right)\left[P_{rot}^{(3)}\left(\omega,\tau\right)\right]^{*} + P_{B}^{(3)}\left(\omega,\Omega_{g},\tau\right)\left[P_{rot}^{(3)}\left(\omega,\tau\right)\right]^{*}\right\} \\
&+ \sum_{g,g'}\left\{P_{A}^{(3)}\left(\omega,\Omega_{g},\tau\right)\left[P_{B}^{(3)}\left(\omega,\Omega_{g'},\tau\right)\right]^{*} + P_{B}^{(3)}\left(\omega,\Omega_{g},\tau\right)\left[P_{A}^{(3)}\left(\omega,\Omega_{g'},\tau\right)\right]^{*}\right\} \\
&+ c.c.
\end{aligned}
\tag{5}
$$

The first term in Eq. (5), as well as those cases of the second term where $g = g'$ do not show any intensity oscillations as function of probe delay.

The second term of Eq. (5) contains terms where $g \neq g'$. Such terms then take a form:

$$
I_{2}\left(\omega,\tau\right) \propto \sum_{g\neq g'}\left\{S_{3}\left(\omega-\Omega_{g}\right)S_{3}^{*}\left(\omega-\Omega_{g'}\right)e^{-i\left[\Omega_{g}-\Omega_{g'}\right]\tau} + S_{3}\left(\omega+\Omega_{g}\right)S_{3}^{*}\left(\omega+\Omega_{g'}\right)e^{i\left[\Omega_{g}-\Omega_{g'}\right]\tau}\right\} \\
+ c.c.
$$

This expression shows oscillations at $(\Omega_{g} - \Omega_{g'})$, the difference between vibrational modes, and the amplitude of the oscillations depends on the optical detection frequency $\omega$.

For a pair of states with energies $\Omega_{g}$ and $\Omega_{g'}$ this combination frequency oscillation will have its maximal amplitude at the maximum of spectral overlap of the two fields: $S_{3}\left(\omega-\Omega_{g}\right)S_{3}^{*}\left(\omega-\Omega_{g'}\right)$ and $S_{3}\left(\omega+\Omega_{g}\right)S_{3}^{*}\left(\omega+\Omega_{g'}\right)$. For pulses of a symmetric Gaussian spectrum, these maxima occur at $\omega = \omega_{3} \pm \dfrac{\Omega_{g}+\Omega_{g'}}{2}$.

The third term in Eq. (5) is a result of interference between vibrational and rotational contributions to nonlinear polarization, and therefore shows intensity oscillations at the fundamental molecular vibrational frequencies:

$$I_3(\omega,\tau) \propto \sum_g \left\{ S_3(\omega-\Omega_g)S_3^*(\omega)e^{-i\Omega_g\tau} + S_3(\omega+\Omega_g)S_3^*(\omega)e^{i\Omega_g\tau} \right\} + c.c.$$

Along the lines of the previous analysis, the amplitude of the intensity oscillations at $\Omega_g$ would be maximized at $\omega = \omega_3 \pm \dfrac{\Omega_g}{2}$, the best spectral overlap between the two contributions.

The fourth term of Eq. (5) indicates interference between two different coherence pathways. Since the contributions of these coherence pathways are symmetric around the probe carrier frequency, the spectral overlap will be maximized around zero detuning. In addition, the sum of the two pathways gives rise to sums of fundamental frequencies, including doubled fundamental frequencies.

$$I_4(\omega,\tau) \propto \sum_{g \neq g'} \left\{ S_3(\omega-\Omega_g)S_3^*(\omega+\Omega_{g'})e^{-i\left[\Omega_g+\Omega_{g'}\right]\tau} + S_3(\omega+\Omega_g)S_3^*(\omega-\Omega_{g'})e^{i\left[\Omega_g+\Omega_{g'}\right]\tau} \right\}$$
$$+ c.c.$$

To summarize this brief theoretical analysis, in TR-FWM in the degenerate case one expects to find the following:

1.  Oscillations at the fundamental frequencies would be most observable at detection frequency detuned (to the blue or to the red) from the carrier by half of a vibrational quantum.

2.  Oscillations at difference of fundamental vibrational frequencies will be maximal at a signal frequency that is detuned from the carrier by the average of the fundamental modes.

3.    Oscillations at sums and doubles of the fundamental vibrational frequencies would be observed at the vicinity of zero detuning from the carrier frequency.

Thus, an operational procedure for analyzing a measured spectrogram is to work in the reverse order of points 1-3 above: first search for high lying features near the zero detuning center line – these will provide clues as to the possible doubled fundamental frequencies or their sums (note that due to the finite available input bandwidth, not all such combinations are certain to appear). Once discovered, arithmetically extract suspected fundamental frequencies, and verify their identity by checking the detuning of their maximal amplitude, and the existence of difference frequencies which may be identified by the signature of their large detuning.

The advantages of the Time Frequency Domain (TFD) data presented above point to the utility of good methods to obtain such data, but as indicated, the acquisition of a spectrum like that in Fig. 1, takes a long time. We now present an attractive alternative way to obtain the two dimensional spectrally resolved information rapidly and without mechanical scanning, a method that is based on Phase Matching Filtering (PMF). Consider the now standard configuration of forward propagating three dimensional (Boxcars) geometry[11,15] where the three input beams constitute the three edges of a square pyramid, and the generated signal completes the fourth edge on the other side of the interaction region (Fig. 2(a)).

Recently we have reported the use of collimated (unfocused) beams in this FWM arrangement for single shot time resolved measurements[16], where the different arrival times of the pulses to different points in the beams' intersection region map the time delays between the pulses. Direct imaging of the signal emerging from each point in the intersection resulted in a single shot picture of several picoseconds of vibrational motion of the molecules captured by a single ultrashort laser pulse. In what follows, we show that by tuning the angle of one of the beams away from a symmetric square pyramid we are able to tune the FWM phase matching frequency, thus tuning the observed generated frequency within the wide spectrum available form the ultrashort pulses. The implemented technique is conceptually akin to well known pulse characterization method of GRENOUILLE[17], that employs the strict phase matching for spectral resolution of nonlinear signals.

The coherent generation of a FWM signal fulfills two conservation criteria: conservation of energy

$$\omega_s = \omega_1 - \omega_2 + \omega_3,\tag{6}$$

and conservation of momentum or phase matching:

$$\left[\omega_1 \hat{k}_1 n(\omega_1) - \omega_2 \hat{k}_2 n(\omega_2) + \omega_3 \hat{k}_3 n(\omega_3)\right]\Big/c = \omega_s \hat{k}_s\, n(\omega_s)\big/c\,.\tag{7}$$

Here $\omega_i$, and $\hat{k}_i$ represent the frequency and direction of propagation for each of the beams and $n(\omega)$ is the refractive index of the medium at a particular frequency. For transparent media one may assume for simplicity a refractive index that is constant over the entire relevant spectral range.

In most experiments the beams are focused into the interaction region, such that each beam contains a range of propagation directions. This virtually relaxes the phase-matching constraints, such that all energetically allowed combinations of frequency components are contributing to the final signal. Assuming Gaussian spectral shape for the incident pulses of the form: $S(\omega) = \varepsilon e^{-\frac{(\omega - \omega_0)^2}{\sigma^2}}$ we obtain the spectral contents of the generated signal:

$$E_s(\omega_s) = \int\limits_{-\infty}^{\infty}\int\limits_{-\infty}^{\infty} d\omega_1 d\Omega \, S(\omega_1) S^*(\omega_2 = \omega_1 - \Omega) S(\omega_3 = \omega_s - \Omega)$$

$$= \frac{\pi\varepsilon^3\sigma^2}{\sqrt{3}} e^{\frac{(\omega_s - \omega_0)^2}{(\sigma\sqrt{3})^2}}. \tag{8}$$

Here $E_s(\omega_s)$ is the amplitude of the generated field at a frequency $\omega_s$, $\varepsilon$ the amplitude of the laser pulse, $\sigma$ is the pulse's bandwidth and $\omega_0$ is its carrier frequency.

Because of the combinations of input frequencies leading to the generated signal, the bandwidth of the generated signal is by a factor of $\sqrt{3}$ larger than that of the fundamental pulse.

The geometrical arrangement of our experiment (Fig. 2(a)) involves unfocused, well collimated beams for which the propagation direction is very well defined[16]. The intersection of the collimated beams is spatially long (several cm), and thus it defines strict phase matching constrains on the interaction between the spectral components of the pulses (for recent detailed discussions of this subject, see Romanov et al.[18]. and Belabes et al.[19]). This tight phase matching may be utilized for Phase Matching Filtering (PMF) for spectrally resolving the FWM signal.

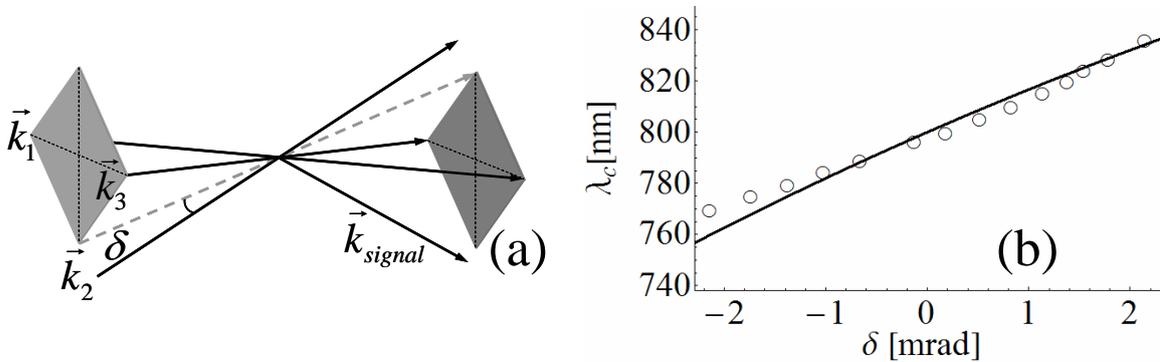

FIG. 2. (a) The 3D-Boxcars configuration, with the Stokes beam deviation angle $\delta$. (b) Central frequency of PMF as function of $\delta$ measured (open circles) and calculated (line).

Figure 2(a) sketches the 3-D folded Boxcars geometry where the propagation of Stokes ($\mathbf{k_2}$) beam is detuned by an angle $\delta$ from the symmetric (square) configuration. The unit propagation vectors of each one of the beams are defined as follows:

$$\vec{k}_1 = \begin{pmatrix} \cos\theta \\ \sin\theta \\ 0 \end{pmatrix}; \ \vec{k}_2 = \begin{pmatrix} \cos\varphi \\ 0 \\ \sin\varphi \end{pmatrix}; \ \vec{k}_3 = \begin{pmatrix} \cos\theta \\ -\sin\theta \\ 0 \end{pmatrix}. \tag{9}$$

Where $2\theta$ is the angle between pump and probe beams ($\mathbf{k_1}$ and $\mathbf{k_3}$), and $\varphi = \theta + \delta$.

For convenience, define $\Omega$ and $\zeta$ :

$$\begin{aligned} \Omega &\equiv \omega_1 - \omega_2 \\ \zeta &\equiv \omega_s - \omega_1 \end{aligned} \tag{10}$$

Using Eq. (6) $\omega_3 = \omega_s - \Omega$.

Substituting Eq. (10) into Eq. (7) we obtain an equation for $\zeta$ :

$$\left[\omega_s - \zeta\right]\hat{k}_1 - \left[\omega_s - \zeta - \Omega\right]\hat{k}_2 + \left[\omega_s - \Omega\right]\hat{k}_3 = \omega_s \hat{k}_s. \tag{11}$$

For pulses which are not "too short" ($\sigma \ll \omega_0$) we use the relation $\Omega \ll \omega_s$ and expand in a power series of $\Omega$ . To the first order we derive a solution for $\zeta$

$$\zeta = a(\theta, \delta)\Omega + b(\theta, \delta)\omega_s. \tag{12}$$

Where the coefficients $a(\theta, \delta)$ and $b(\theta, \delta)$ depend only on the geometry of the experiment.

$$\begin{aligned} a &= \frac{\cos\delta - 2\cos(2\theta) + \cos(2\theta + \delta)}{-2 + \cos\delta + \cos(2\theta + \delta)} - \frac{4\sin^2\theta}{\cos\delta - 2\cos(2\theta) + \cos(2\theta + \delta)}; \\ b &= \frac{4\cos\theta\left(\cos(\theta + \delta) - \cos\theta\right)}{-2 + \cos\delta + \cos(2\theta + \delta)}. \end{aligned} \tag{13}$$

Next we calculate the spectral characteristics of the FWM signal by superposing all allowed combinations of spectral components of the input pulses:

$$E_{PM}\left(\omega_s;\theta,\delta\right) = \int_{-\infty}^{\infty} d\Omega S\left(\omega_s - \zeta\right) S^*\left(\omega_s - \zeta - \Omega\right) S\left(\omega_s - \Omega\right)$$

Here $E_{PM}\left(\omega_s;\theta,\delta\right)$ is the phase-matched signal field amplitude for a particular choice of the angles between incident beams.

Using the same spectral form of the laser pulses as before, the phase matched FWM spectrum may be represented as:

$$E_{PM}\left(\omega_s;\theta,\delta\right) \propto e^{-\frac{4\ln 2\left(\omega_s - \omega_c\right)^2}{\left(\Delta\omega\right)^2}}.$$

Here $\omega_c$ is the central frequency of the emitted field and $\Delta\omega$ is its Full Width at Half Maximum.

Naturally both central frequency and the bandwidth depend on the geometry:

$$\omega_c\left(\theta,\delta\right) = 2\frac{\left[a^2 - a\left(1+b\right)+1+b\right]}{2+2a^2+4b+3b^2-2a\left(1+2b\right)}\omega_0;$$

$$\Delta\omega_s\left(\theta,\delta\right) = 2\sigma\sqrt{\log 2}\sqrt{\frac{2\left(1+a+a^2\right)}{\left(2+2a^2+4b+3b^2\right)-a\left(2+4b\right)}}$$

Where the coefficients $a$ and $b$ are given by Eq. (13). The second square root in the expression for $\Delta\omega_s$ provides a "narrowing factor", defining the ratio between the bandwidth of the incident pulse and this of the generated FWM.

As a test case, consider this result for square Boxcars configuration, namely for $\delta = 0$. For this case we obtain: $\omega_c\left(\theta,0\right) = \omega_0$ and $\Delta\omega_s\left(\theta,0\right) = 2\sigma\sqrt{\log\left(2\right)}\big/\sqrt{3}$. As could be expected for this symmetric configuration, the central frequency of the emitted signal coincides with the carrier frequency of the laser, but the bandwidth is by a factor of $\sqrt{3}$ **narrower** than that of the laser pulse, and thus **by a factor of 3 narrower** than that of the signal generated by focused beams (see Eq. (8) above), indicating that the emitted signal is effectively filtered by the phase-matching requirement.

A convenient expression may be obtained for small deviations of the Stokes beam from the square Boxcars geometry ($\delta \ll \theta$)

In such a case

$$\omega_c\left(\theta, \delta\right) \simeq \omega_0 \left( 1 - \frac{4}{3} \delta \cot \theta \right). \tag{14}$$

Eq. (14) shows that for small deviations, the detuning of the Phase Matching Filter (PMF) from the laser frequency is linear with the angular deviation of the Stokes beam. This derived analytical result is plotted as a solid line in Fig. 2(b). The simple dependence provides a convenient means for spectral scanning of the frequency of the phase matched FWM signal.

We now combine the PMF with the single shot method for measuring spectrally AND temporally resolved FWM in chloroform. The experimental conditions are as follows: the cross section of the collimated incident beams is 5mm, the angle between pump and probe beam is 8 degrees. Neat spectroscopic grade chloroform was used as purchased from Sigma-Aldrich. The sample was placed in a 35mm long spectroscopic cell. The laser pulses are derived from Spectra-Physics Oscillator (Tsunami) and amplified in a Spectra-Physics Chirped pulse amplifier (Spitfire). The pulses are 60 fs long, are centered around 800 nm and of a bandwidth of 400cm[-1], and their energy at the entrance to the cell is 1 mJ per pulse. The FWM beam is imaged on a CCD camera, ISG-1394-S CMOS camera. To ensure that every image is indeed produced by single laser pulse, the repetition rate of the laser is reduced to 10Hz, and the exposure of the camera is set to 40ms. The angular deviation of the Stokes beam is achieved by fine angular rotation of the last folding mirror before the cell, and is measured as a displacement of this beam on a target located 4m away. The spectral calibration of each set of experiments is performed by measuring the coherence spike (signal generated at the coincidence of all three pulses) with JY TRIAX 180 monochromator coupled to CCD camera.

By tuning the angle $\delta$ we obtain a series of time resolved images, each measured within a single laser shot. An example of such an image is shown in (Fig. 3(a)) for the angle $\delta = 0$, $\Delta_c = 0$, the retrieved temporal profile is shown at the inset of the panel. Measurements like the one depicted in Fig. 3(a) are repeated for a range of angles, each such temporal profile is Fourier Transformed, to produce a line in the two dimensional map (Fig. 3(b)) of the power spectrum (vertical axis) of the FWM signal

The data included in Fig. 1(b) and Fig. 3(b) are in principle equivalent; the spectral and temporal resolutions are parameters of choice, determined by the spectrometer resolution on the one hand, and pulse duration and the number of measured angles on the other. The experimental simplicity of the single shot PMF measuring method offers significant advantages: no spectrometers and no critical alignment, simply broad unfocused beams crossing in the sample. Moreover, since there are no mechanical scans of time delays, and since the entire measurement may be derived within a single ultrafast pulse per spectral point, a full measurement may be performed with very few laser pulses – a crucial requirement for molecules undergoing rapid photo-bleaching as is the case for many molecules of biological interest[20].

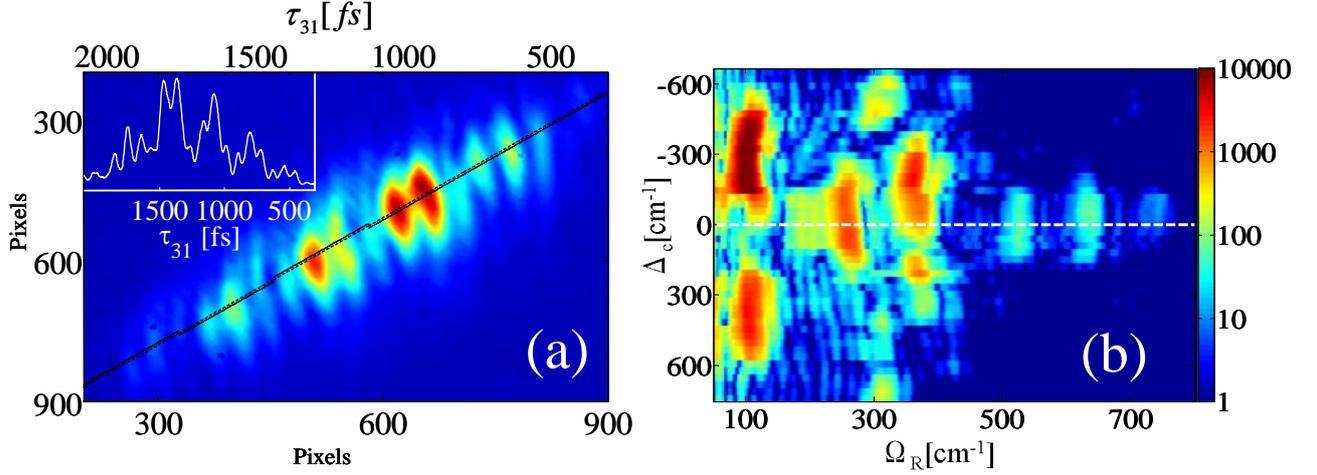

FIG. 3. (Color online) (a) Typical Single Shot FWM image of chloroform with the time resoled FWM signal in the inset. (b) Two dimensional spectrogram of the chloroform Raman FWM signal obtained by phase matching filtering, where the power spectrum of the FWM signal is plotted horizontally against the Raman frequency ($\Omega_R$) and vertically against ($\Delta_c$), the detuning of the PMF from the laser center frequency (log scale).

Consider the distinct spectral features of the spectrogram shown in Fig. 3(b):

There are several distinct spectral features at: 104, 261, 300, 365, 520, 625 and 730 cm$^{-1}$. The bandwidth of our pulses does not allow the observations of features at higher frequencies. Figure 4 depicts, for several of these features, vertical cuts in Fig. 1(b) and Fig. 3(b), showing the intensity distribution at a given Raman frequency $\Omega_R$ as a function of the detuning from the laser center frequency $\Delta_c$. In the Fig. 4, the spectral profiles are given for both measurement methods, those obtained by acquiring a full spectrum at a given delay (Fig. 1(b)), and by the newly introduced method of Single-Shot PMF (Fig. 3(b)).

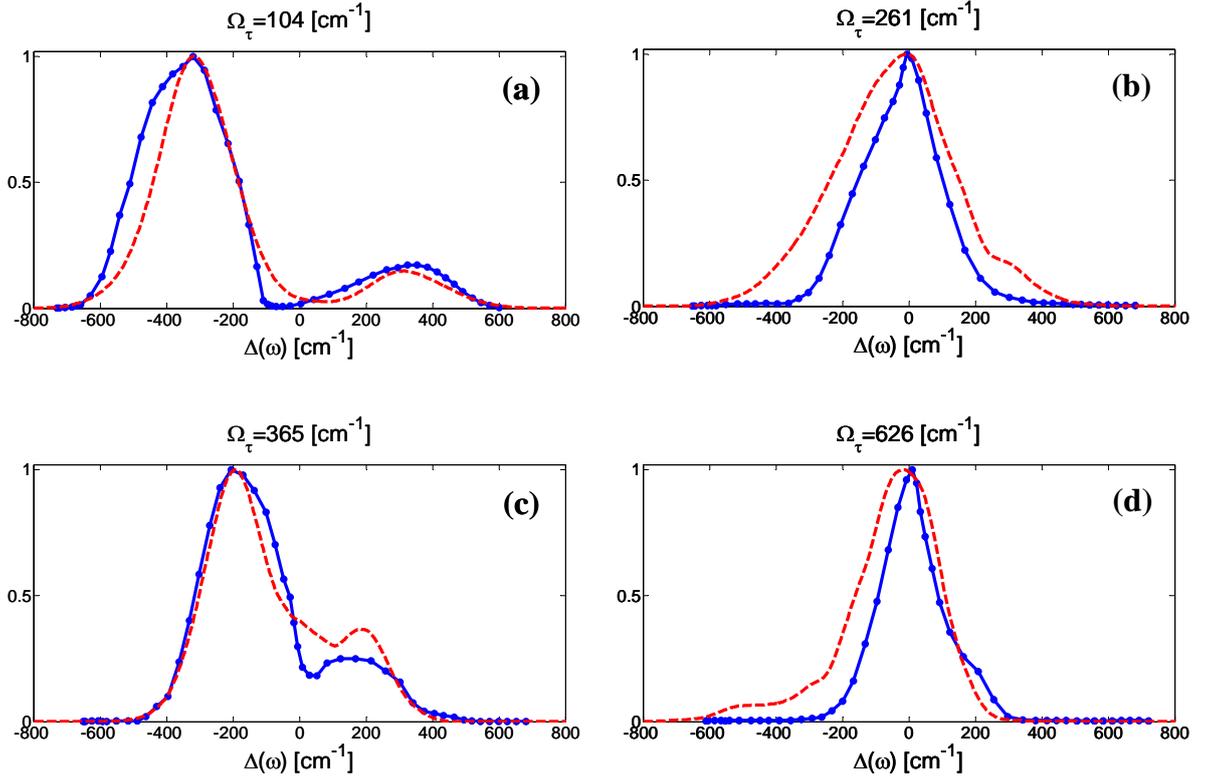

FIG. 4. (Color online) Detuning dependence (normalized) of the Raman contributions and a comparison between the 2D-spectrograms obtained by the Spectrally Resolved Temporal Scanning method (red dashed lines) and by the Single-Shot PMF method (blue lines with solid circles). (a) Depicts section of both spectrograms along $\Omega_R = 104 cm^{-1}$ (b) along the $\Omega_R = 261 cm^{-1}$ (c) $\Omega_R = 365 cm^{-1}$ and (d) $\Omega_R = 626 cm^{-1}$.

The known fundamental mode frequencies of chloroform are given in Table II.

**Table II.** The fundamental modes Vibrational frequencies of chloroform.

| Vibration | Character | Energy(cm$^{-1}$) |
|---|---|---|
| $\nu_1$ | C-H $s$ stretch | 3021 |
| $\nu_2$ | CCl$_3$ $s$ stretch | 680 |
| $\nu_3$ | CCl$_3$ $s$ deform | 365 |
| $\nu_4$ | C-H $d$ bend | 1220 |
| $\nu_5$ | CCl$_3$ $d$ stretch | 774 |
| $\nu_6$ | CCl$_3$ $d$ deform | 261 |

The results obtained by both methods are quite similar, and will be discussed together. The spectral profile of both $\Omega_\tau = 104\,cm^{-1}$ and $\Omega_\tau = 365\,cm^{-1}$ show double maxima, peaking at $\Delta(\omega) = \pm 310\,cm^{-1}\left[\simeq (261+365)/2\right]$ and $\Delta(\omega) = \pm 183\,cm^{-1}\left[\simeq 365/2\right]$ respectively. The spectral profile of the $\Omega_\tau = 261\,cm^{-1}$ line should have shown a double maximum at $\Delta(\omega) = \pm 130\,cm^{-1}\left[\simeq 261/2\right]$, but these are unresolved as they are measured with our 400cm$^{-1}$ pulse. The peaks at 522 cm$^{-1}$ (=261x2), 730 cm$^{-1}$(=365x2) and 626 cm$^{-1}$ (=261+365) are centered at $\Delta(\omega) = 0$ as theoretically predicted for interference between contributions of two spectroscopic pathways. The fact that the double peak lines are asymmetric in their intensity is puzzling. It seams to be a result of imperfect compensation of spectral phases of the laser pulse, as evident from the curvature of the fringes seen in Fig. 1(a), but this observation requires further experiments and analysis.

In conclusion, we have discussed in detail the advantages offered by working in the combined Time-Frequency domain. We demonstrated that while the basic spectral information (i.e. the measured frequencies) may be derived from simple time resolved measurements, the combined TFD measurements provide unequivocal characterization of the individual lines and identifies them as either fundamental modes or beat frequencies between such modes. We further showed that in time resolved FWM measurements, a second Feynman diagram, not normally included in the analysis of such experiments, must be considered for proper interpretation of the data. These results, and the additional insight as to the analysis of the observed spectra are obtainable in 'standard' Time-Frequency measurements, but as is well known, these are long measurements where for each delay, a spectrogram needs to be measured, later to be Fourier transformed to yield data of the type presented in Fig. 1(b). Long measurements are problematic for molecules susceptible to photobleaching, or other light induced damage, a concern afflicting many molecules of biological interest where detailed dynamic and structural studies are desired. In this paper we have introduced and demonstrated a new approach to TRFWM, where the entire time domain signal is captures within a single pulse. By taking advantage of the unique geometrical arrangement, we use Phase Matching Filtering (PMF), where the collimated beams impose well defined phase matching conditions on the intersecting beams, such that only a narrow-frequency section of the spectrally broad pulses is being phase matched, and therefore it is only this relatively narrow frequency range that contributes to the FWM signal. We demonstrated the method experimentally, and performed detailed comparison to data obtained by the traditional manner. The Single-Shot PMF method offers significant advantages, being fast and efficient in terms of the number of laser pulses used for the entire measurement, and experimentally simple. Because of its inherent speed, the new method has a promising potential for the characterization of short lived optically unstable molecules.

Experiments are under way with more complex molecules, and for the implementation of the method for studying transient excited electronic states.

We gratefully acknowledge discussions with Ilya Averbukh, Sharly Fleischer and Mark Vilensky. This work was supported by the Nancy and Steve Grand Center for Sensors and Security, by a James Franck Minerva grant, and by the Israel Science Foundation.

[*] These authors made an equal contribution to this work

[1] D. M. Jonas, Annu. Rev. Phys. Chem. **54**, 425 (2003).

[2] D. Keusters, H. S. Tan, and W. S. Warren, J. Phys. Chem. A **103**, 10369 (1999).

[3] M. H. Levit, *Spin Dynamics* (John Willey & Sons, Ltd., Chichester, 2001).

[4] S. Mukamel, *Principles of Nonlinear Optical Spectroscopy* (Oxford University Press, New York, 1995).

[5] M. Heid, S. Schlucker, U. Schmitt, T. Chen, R. Schweitzer-Stenner, V. Engel, and W. Kiefer, J. Raman Spectrosc. **32**, 771 (2001).

[6] B. D. Prince, A. Chakraborty, B. M. Prince, and H. U. Stauffer, J. Chem. Phys. **125**, 044502 (2006).

[7] S. Nath, D. C. Urbanek, S. J. Kern, and M. A. Berg, Phys. Rev. Lett. **97**, 267401 (2006).

[8] S. Nath, D. C. Urbanek, S. J. Kern, and M. A. Berg, J. Chem. Phys. **127**, 044307 (2007).

[9] S. O. Konorov, X. G. Xu, R. F. B. Turner, M. W. Blades, J. W. Hepburn, and V. Milner, Opt. Express **15**, 7564 (2007).

[10] X. J. G. Xu, S. O. Konorov, S. Zhdanovich, J. W. Hepburn, and V. Milner, J. Chem. Phys. **126**, 091102 (2007).

[11] Y. Prior, Appl. Opt. **19**, 1741 (1980).

[12] D. C. Urbanek and M. A. Berg, J. Chem. Phys. **127**, 044307 (2007).

[13] Y. Prior, IEEE J. Quantum Electron. **20**, 37 (1984).

[14] A. Shalit, Y. Paskover, and Y. Prior, Chem. Phys. Lett. **450**, 408 (2008).

[15] J. A. Shirley, R. J. Hall, and A. C. Eckbreth, Opt. Lett. **5**, 380 (1980).

[16] Y. Paskover, I. S. Averbukh, and Y. Prior, Opt. Express **15**, 1700 (2007).

[17] P. O'Shea, M. Kimmel, X. Gu, and R. Trebino, Opt. Lett. **26**, 932 (2001).

[18]D. Romanov, A. Filin, R. Compton, and R. Levis, Optics Letters **32**, 3161 (2007).

[19]N. Belabas and D. M. Jonas, J. Opt. Soc. Am. B **22**, 655 (2005).

[20]H. Kawano, Y. Nabekawa, A. Suda, Y. Oishi, H. Mizuno, A. Miyawaki, and K. Midorikawa, Biochem. Biophys. Res. Commun. **311**, 592 (2003).